\documentclass[copyright,creativecommons]{eptcs}
\providecommand{\event}{QAPL 2014} 
\usepackage{breakurl}              

\usepackage{array}
\usepackage{amsfonts}
\usepackage{amssymb}
\usepackage{amsmath}
\usepackage{amsthm}
\usepackage[crop=off, off]{auto-pst-pdf}

\usepackage{caption}
\usepackage{subfig}
\usepackage{pepa}
\usepackage{extarrows}
\usepackage{enumerate}
\usepackage{dsfont}
\usepackage{todonotes}
\usepackage{graphicx}
\usepackage{url}

\newtheorem{definition}{Definition}

\newtheorem{theorem}{Theorem}

\newtheorem{remark}{Remark}
\newtheorem{example}{Example}

 \newcommand{\xe}{\mathrm{e}}            
 \newcommand{\xR}{\mathbb{R}}
 
 \newcommand{\norm}[1]{%
   \left\lVert #1 \right\rVert%
    }

\newcommand{\pepa}[1]{\ensuremath{\mathit{#1}}}
\newcommand{\dsl}{\displaystyle}
\newcommand{\crate}[1][\alpha]{\mathcal{R}_{#1}}    
\newcommand{\arate}[1][\alpha]{r_{#1}}              
\newcommand{\act}[1]{\xlongrightarrow{#1}}          

\begin{document}

\title{Extended Differential Aggregations in Process Algebra for Performance and Biology}
\author{Max Tschaikowski
\institute{University of Southampton \\ United Kingdom}
\email{m.tschaikowski@soton.ac.uk}
\and
Mirco Tribastone
\institute{University of Southampton \\ United Kingdom}
\email{m.tribastone@soton.ac.uk}
}

\def\titlerunning{Extended Differential Aggregations in Process Algebra for Performance and Biology}
\def\authorrunning{M. Tschaikowski \& M. Tribastone}

%
%

\allowdisplaybreaks[3]
\maketitle

\begin{abstract}
We study aggregations for ordinary differential equations induced by fluid semantics for Markovian process algebra which can capture the dynamics of performance models and chemical reaction networks. Whilst previous work has required perfect symmetry for exact aggregation, we present \emph{approximate fluid lumpability}, which makes nearby processes perfectly symmetric after a perturbation of their parameters. We prove that small perturbations yield nearby differential trajectories. Numerically, we show that many heterogeneous processes can be aggregated with negligible errors.
\end{abstract}

\section{Introduction}\label{sec:intro}

%

Fluid semantics for process algebra describe the dynamics of a model in terms of a system of ordinary differential equations (ODEs), which can be interpreted as a deterministic approximation to the expectation of the stochastic process underlying the classical Markovian semantics (e.g., \cite{Hillston_QEST05,Cardelli_TCS_08,Ciocchetta_Hillston_09,HaydenLipschitz,DBLP:journals/tse/TribastoneGH12}). When the model under consideration consists of many copies of processes in parallel, the ODE system size is independent from the multiplicities of such copies, unlike the Markovian representation that is well-known to suffer from state explosion.

Unfortunately, not every process algebra model enjoys a compact ODE description.
A possible solution to this problem is to exploit symmetries, captured by appropriate behavioural relations, that give rise to an \emph{aggregated} ODE system whose solution can be related to that of the original, potentially massive, one. Arguably, in the process algebra literature this topic has so far received less attention than its stochastic counterpart, concerned with developing behavioural equivalences that that induce state-space aggregations in the underlying Markov chain (e.g.,~\cite{DBLP:conf/esm/HermannsR98,canonical,bucholz94papm}). Indeed, in~\cite{concur12,efl_tcs} we studied a notion of behavioural equivalence for ODEs in the context of the stochastic process algebra PEPA~\cite{pepa}; this has found application to reducing the ODE system sizes of hierarchical models exhibiting replicas of massively parallel composite processes with arbitrary levels of nesting~\cite{Tschaikowski2013a}. ODE aggregations have also been investigated in~\cite{DBLP:conf/lics/DanosFFHK10} for rule-based models such as Kappa~\cite{Danos200469} and BioNetGen~\cite{Faeder:2009aa}, for the modelling and analysis of biomolecular networks.

The goal of this paper is to extend the toolkit of ODE aggregations available for stochastic process algebra, by making the following contributions.

\paragraph{\bf A unified framework for performance and biological modelling.}   Let us first observe that PEPA and rule-based models have complementary domain-specific semantics of synchronisation that inevitably circumscribe the scope of validity of the results on ODE aggregations. Indeed, while PEPA can be particularly useful for the performance evaluation of computing systems, rule-based models use the \emph{law of mass action}. This is well known to be at the basis of biochemical reaction networks, though it has also been employed in epidemiological models (e.g.,~\cite{Watson1972}) as well as in certain wireless networks (e.g.~\cite{Wang22052009}).

Our initial starting point is to consider a unified framework that encompasses both kinds of interaction.  Specifically, we introduce and study \emph{Fluid Extended Process Algebra} (FEPA), a lightweight and conservative extension of \emph{Fluid Process Algebra}  presented in \cite{concur12}, featuring a more general parallel operator that can capture both dynamics.  We will first show that the notion of \emph{exact fluid lumpability} (EFL) of~\cite{concur12} carries over to FEPA\@. This will be used to set the stage for approximate notions of ODE aggregations, discussed later in the paper, which are defined in terms of their exact counterparts.

For an informal overview of our results, let us consider the process
\begin{equation}\label{eq:example}
\Big( P_1[N_1] \parallel_K P_2[N_2] \parallel_K \cdots \parallel_K P_D[N_D] \Big) \parallel_{L} Q[M]
\end{equation}
where, for all $1 \leq i \leq D$, $P_i$ is some sequential component that is replicated $N_i$ times, and $\parallel_{K}$ is the (generic) parallel operator, parameterised by an action set $K$, in a CSP-like fashion. EFL may essentially reduce the analysis of such a model by considering the fluid trajectory of a \emph{representative} $P_i$, which is shown to be equal to that of any other $P_j$ if $N_i = N_j$ and  $P_i$ and $P_j$ are isomorphic. Thus, symmetry is required both at the level of the sequential component and at the compositional level, by ensuring that all populations have the same size.

%
%
\paragraph{\bf Ordinary fluid lumpability.} We relax the symmetry requirements of EFL by introducing the notion of \emph{ordinary fluid lumpability} (OFL). Similarly to EFL, it considers symmetry through isomorphism at the sequential level. However, it allows \emph{heterogeneity} at the compositional level: in the example above, it may yield an exactly aggregated ODE system even if $N_i \neq N_j$.  However, unlike EFL, where all the trajectories of the original ODE system can be obtained from the solution of the aggregate, in OFL the aggregate gives the exact sum of the solutions of its parts, but their individual trajectories cannot be recovered. In this sense, it corresponds to the aggregation presented in~\cite{DBLP:conf/lics/DanosFFHK10}, but with some differences. In~\cite{DBLP:conf/lics/DanosFFHK10} aggregates can be obtained by collapsing non-isomorphic biochemical species. This is useful in that application domain, where species are modelled as non-atomic entities with an internal state characterised by the simplest agents of which they are formed. This is in contrast to the modelling scenario envisaged in our process algebra, where atomic entities do not combine, but only interact with other atomic entities. More specifically, a typical modelling pattern amenable to our notions of aggregation is that of \emph{multi-class} systems, where analogous processes (e.g., two or more kinds of infections~\cite{Watson1972}) exhibit similar behaviour, but with different rates and/or with different multiplicites. Finally, unlike~\cite{DBLP:conf/lics/DanosFFHK10}, we also focus  on compositionality properties of our aggregations.

%

\paragraph{\bf Approximate aggregations.}
We aim to go beyond \cite{concur12}, \cite{DBLP:conf/lics/DanosFFHK10}, as well as OFL, in that we also relax the requirement on the \emph{exactness} of the aggregation. This fact partly stems from the use of strong symmetry at the sequential level. However, there is evidence of criticism on this assumption when models confront real systems, where the difficulty in measuring rates may induce estimation errors that numerically tell apart apparently  identical agents (e.g.,~\cite{springerlink:10.1007/BFb0039071,1029849}). This has motivated work on approximate reasoning with probabilistic and stochastic models (e.g.,~\cite{worrell01:approximate,DBLP:conf/concur/PierroHW03}).
In this paper, we study $\varepsilon$-variants of both EFL and OFL as a means to relaxing symmetries at the sequential level. These variants allow non-isomorphic processes to be aggregated if there exists a \emph{perturbation} in the rates make them isomorphic. For instance, let us take
$P_i \xrightarrow{(\alpha, r)} P_k$ and $P_j \xrightarrow{(\alpha, r + \varepsilon)} P_k$,
   for some $P_k$, where the edges give the usual action/rate pairs, with $r > 0$ and $\varepsilon > 0$. Then, whilst these processes cannot be aggregated with either EFL or OFL, it holds for example that they are $\varepsilon$-ordinarily fluid lumpable for any $N_i$ and $N_j$. Clearly, the aggregated system will not be in exact correspondence with the original one. However, we provide a theoretical bound that shows that the aggregation error  depends \emph{linearly} in the intensity of the perturbation $| \varepsilon |$.

Exhibiting such near-symmetries may appear quite limiting for practical applications; however, there are models in the literature that do exhibit this characteristic  (see, e.g.,~\cite{dsn2013} and references therein). When such a condition is satisfied, it is possible to systematically construct a reduced model independently from the abundances of the species involved, or from the speed at which certain interactions occur, as is required for approximate aggregation methods based on quasi steady-state or quasi-equilibrium (see, e.g.,~\cite{10.3389/fgene.2012.00131}). In this respect, this paper is more closely related to~\cite{dsn2013}, which studies how perturbations of model parameters can lead to ODE aggregations. However, while \cite{dsn2013} is more general in that the ODEs are not necessarily derived from a process algebra, it is  restricted to aggregations of OFL type. Furthermore, \cite{dsn2013} does not address compositionality issues, i.e., how to reuse ODE aggregations of a model in another context.
   %

 \paragraph{\bf Characterisation of ODE aggregations.} We characterise the nature of such aggregations in two main ways. (i) Firstly, we show that all our ODE aggregations can be induced by suitable notions of behavioural equivalence, which turn out to be congruences for FEPA\@.
 (ii) Secondly, we provide some numerical evidence on the usefulness of the approximate versions of EFL and OFL, presenting model examples where even significant perturbation in the rates may yield negligible error in practice.

\section{Fluid Extended Process Algebra}

To define FEPA, we exploit the fact that fluid semantics reason about representatives of replicated sequential components, which we will also call \emph{fluid atoms}. For instance, in~(\ref{eq:example}), $P_1[N_1]$ will be represented in FEPA using a single fluid atom $P_1$ together with the information about the multiplicity of replicas, encoded in a \emph{population function}. Let us first define the grammar for fluid atoms.
\begin{definition}\label{def:fepa}
The syntax of a FEPA fluid atom is given by
\begin{align*}
S & ::= (\alpha, r).S \mid S + S \mid A  & \text{with} \quad A \rmdef S \text{\quad(constant)},
\end{align*}
where $\alpha$ is an action in the \emph{action set} $\mathcal{A}$  and $r \in \mathds{R}_{>0}$.
\end{definition}
The structured operational semantics is given by the following four rules:
\[
\frac{\phantom{P \act{(\alpha, r)} P'}}{(\alpha,r).P \act{(\alpha, r)} P} \ ,
\ \
\frac{P \act{(\alpha, r)} P'}{P + Q \act{(\alpha, r)} P'} \ ,
\ \
\frac{Q \act{(\alpha, r)} Q'}{P + Q \act{(\alpha, r)} Q'} \ ,
\ \
\frac{P \act{(\alpha, r)} P'}{A \act{(\alpha, r)} P'} \ A \rmdef P .
\]
Given a fluid atom $P$, these rules induce a labelled transition system, denoted by $dg(P)$ (the derivation graph), with nodes denoted by $ds(P)$ (the \emph{derivative set}), and with a transition multi-set  where transitions have a  multiplicity equal to the number of distinct derivations between any two fluid atoms.
%

Semi-isomorphism, at the basis of our characterisation results, relates two fluid atoms whenever their derivation graphs are isomorphic up to replacing multiple equally-labelled transitions between two states with a single transition with the same action type and the rate sum across all such transitions.

\begin{definition}[Semi-Isomorphism]\label{def_semi-iso}
Two FEPA sequential components $P$ and $Q$ are semi-isomorphic if there is a bijection $\sigma : ds(P) \rightarrow ds(Q)$ which satisfies $\sum_{P_i \xrightarrow{(\alpha, r)} P_j} r $ $= \sum_{\sigma(P_i) \xrightarrow{(\alpha, r)} \sigma(P_j)} r$ for all $P_i, P_j \in ds(P)$ and $\alpha \in \mathcal{A}$. We shall call such a bijection $\sigma$ a semi-isomorphism.
\end{definition}
For instance $P \rmdef (\alpha, r_1).P + (\alpha, r_2).P$ is semi-isomorphic to $Q \rmdef (\alpha, r_1 + r_2).Q$.

A FEPA model is a composition of fluid atoms, using the parallel operator $\parallel_L$. As in~\cite{concur12}, FEPA does not feature the hiding operator.
\begin{definition}[FEPA Model]
A FEPA model $M$ is given by the grammar
\[ M ::= M \parallel_L M \mid P \]
where $L \subseteq \mathcal{A}$ and $P$ is a fluid atom. For any two distinct fluid atoms $P$ and $P'$ in $M$, we require that $ds(P) \cap ds(P') = \emptyset$.
\end{definition}
In comparing derivative sets, equality is intended to be syntactical. The requirement on pairwise disjoint derivative sets is without loss of generality: If two distinct fluid atoms have a common derivative, it is always possible to relabel one atom with appropriate fresh constants so as to satisfy the above condition.

\begin{example}\label{ex:first}
Let us consider the FEPA process
\begin{equation}\label{eq_running_example_fpa}
\pepa{Sys} := (P_1 \parallel_\emptyset \ldots \parallel_\emptyset P_D) \parallel_{\{\alpha\}} Q \ ,
\end{equation}
with fluid atoms given by
\begin{align*}
P_d & \rmdef (\alpha, r).P'_d, & P'_d & \rmdef (\beta, s).P_d,& Q & \rmdef (\alpha, u).Q', & Q' & \rmdef (\delta, w).Q, & 1 \leq d \leq D.
\end{align*}
Intuitively, the above model considers a situation where $D$ \emph{independent} groups of agents, recognisable by the empty cooperation sets, synchronise with a common group of agents, of type $Q$, through action $\alpha$.
\end{example}


\begin{definition}
Let $M$ be a FEPA model. We define then
$\mathcal{G}(M)$ as the set of all fluid atoms in $M$; $\mathcal{B}(M)$ as the set of all atoms' derivatives, i.e., $\mathcal{B}(M) = \bigcup \{ ds(P) \ | \ P \in \mathcal{G}(M) \}$; a \emph{population function} $V : X \rightarrow \mathbb{R}_{\geq 0}$ with $\mathcal{B}(M) \subseteq X$; and an \emph{initial} population function $V(0) : X \rightarrow \mathbb{N}_0$.
\end{definition}
For instance, $\mathcal{G}(\pepa{Sys}) = \{P_1,\ldots,P_D,Q\}$, $\mathcal{B}(\pepa{Sys}) = \{ P_d, P'_d \mid 1 \leq d \leq D \} \cup \{ Q, Q' \}$. Function $V(0)$ encodes the size of the system at time $t = 0$. For instance,
\begin{align}\label{eq:v0}
V_{P_d}(0) & = N_d, & V_{P'_d}(0) & = 0, & V_Q(0) & = N_Q, & V_{Q'}(0) & = 0,
\end{align}
states that at $t = 0$ there are $N_d$ agents in the state $P_d$, no agents in the state $P'_d$, $N_Q$ agents in the state $Q$ and no agents in the state $Q'$.

We are now ready to provide the semantics for interaction in FEPA. We consider two instances, distinguished by the choice of a (binary) \emph{synchronisation function} that is hereafter denoted by $\rho$. Choosing $\rho = \min$ yields the minimum-based semantics of PEPA; the law of mass action is instead recovered by choosing $\rho = \cdot$ (intended as multiplication).  With this latter choice, FEPA can be seen as the fluid counterpart of Markovian process algebra such as~\cite{bucholz94papm}, or as an alternative to process algebra for biological networks such as Bio-PEPA~\cite{Ciocchetta_Hillston_09}.

\begin{definition}[Apparent Rate]\label{def:app.rate}
The apparent rate of action $\alpha$ in a FEPA fluid atom $P$, denoted by $r_\alpha(P)$, is defined as follows:
\[
r_\alpha((\beta,r).S)  :=
\left \{
\begin{array}{lllrl}
r & , \ \beta = \alpha, & \qquad \qquad \qquad \qquad & r_\alpha(A)  & := r_\alpha(S), \  A \rmdef S ,\\
0 & , \ \text{else,} & \qquad \qquad \qquad  \qquad & r_\alpha(P + Q)  & := r_\alpha(P) + r_\alpha(Q).
\end{array}
\right.
 \]
\end{definition}

\begin{definition}[Parameterised Apparent Rate]\label{def:app.rate.v}
Let $M$ be a FEPA model, $\alpha \in \mathcal{A}$, $V$ a population function, and $\rho$ the synchronisation function.
The apparent rate of $M$ with respect to $V$ is defined as
\[
\begin{split}
\arate(M_0 \parallel_L M_1, V) & :=
    \begin{cases}
        \min(\arate(M_0, V), \arate(M_1, V))           & , \ \alpha \in L \ \land \ \rho = \min , \\
        \arate(M_0, V) \cdot \arate(M_1, V)            & , \ \alpha \in L \ \land \ \rho = \cdot , \\
        \arate(M_0, V) + \arate(M_1, V)                & , \ \alpha \notin L.
    \end{cases} \\
\arate(P, V) & := \sum_{P_i \in ds(P)} V_{P_i} \arate(P_i),
\end{split}
\]
where $\arate(P_i)$ is the apparent rate of a FEPA fluid atom $P_i$, by Definition~\ref{def:app.rate}.
\end{definition}
For instance, in (\ref{eq_running_example_fpa}) it holds that $\arate(P_d, V) = r V_{P_d}$, which gives the apparent rate at which a population of $V_{P_d}$ $P_d$-components exhibits action $\alpha$. Let us assume that $D = 1$ in (\ref{eq_running_example_fpa}). Then $r_\alpha(P_1 \parallel_{\{\alpha\}} Q, V) = \min (r V_{P_1}, u V_Q)$ if $\rho = \min$; for $\rho = \cdot$, instead, we have that $r_\alpha(P_1 \parallel_{\{\alpha\}} Q, V) = r\cdot u \cdot V_{P_1} \cdot V_Q$. In this case, the model corresponds to a chemical reaction network which may be expressed, using standard notation, by $P_1 + Q \rightarrow P_1' + Q'$, with rate constant equal to $r \cdot u$, and two monomolecular reactions, $P_1' \rightarrow P_1$ and $Q' \rightarrow Q$, with rate constants $s$ and $w$, respectively.

The following quantities are used to define the vector field of the ODE system to be analysed.

\begin{definition}[Parameterised Component Rate]\label{def:rate.v}
Let $M$ be a FEPA model, $\alpha \in \mathcal{A}$ and $V$ a population function. The component rate of $P' \in \mathcal{B}(M)$ is parameterised by $V$ in the following manner.
\begin{itemize}
\item $M = M_0 \parallel_L M_1$: if $P' \in \mathcal{B}(M_i)$, $i = 0,1$, and $\alpha \in L$ then
\[
 \crate(M_0 \parallel_L M_1, V, P') :=
    \dsl \frac{\crate(M_i, V, P')}{\arate(M_i, V)} \arate(M_0 \parallel_L M_1, V).
\]
\item $M = M_0 \parallel_L M_1$: if $P' \in \mathcal{B}(M_i),$ $i = 0,1$, and $\alpha \notin L$ then
\[
 \crate(M_0 \parallel_L M_1, V, P') := \crate(M_i, V, P').
\]
\item $M = P$: then
\[
\crate(P, V, P') := V_{P'} \arate(P') .
\]
\end{itemize}
\end{definition}

\paragraph{Notation.}  We use Newton's dot notation $\dot{V}_P$ for the derivative of $V_P$. To enhance readability, time $t$ will be suppressed, e.g., $\dot{V}_P$ denotes $\dot{V}_P(t)$. 

\begin{definition}\label{def_35}
Let $M$ be a FEPA model. The initial value problem for $M$ is given by $\dot{V} = F(V)$ with initial condition $V(0)$, where
\[ F_P(V) := \sum_{\alpha \in \mathcal{A}} \bigg( \Big( \sum_{P' \in \mathcal{B}(M)} p_\alpha(P', P) \crate(M, V, P') \Big) - \crate(M, V, P) \bigg) \]
and
\[
p_\alpha(P,P') = \frac{1}{\arate(P)} \sum_{P \xrightarrow{(\alpha, r)} P'} r
\]
for all $\alpha \in \mathcal{A}$ and $P,P' \in \mathcal{B}(M)$.
\end{definition}

For instance, the initial value problem of (\ref{eq_running_example_fpa}) and (\ref{eq:v0}) in the case of $\rho = \cdot$ is given by the initial condition (\ref{eq:v0}) and the ODE system
\begin{align}\label{eq:ode}
\dot{V}_{P_d} & = - r u V_{P_d} \cdot V_{Q} + s V_{P'_d} & \dot{V}_{P'_d} & = - s V_{P'_d} + r u V_{P_d} \cdot V_{Q} \nonumber \\
\dot{V}_{Q} & = - u V_{Q} \cdot \sum_{1 \leq d \leq D} r V_{P_d} + w V_{Q'} & \dot{V}_{Q'} & = - w V_{Q'} + u V_{Q} \cdot \sum_{1 \leq d \leq D} r V_{P_d}
\end{align}


The notion of well-posedness given below is needed to characterise ODE aggregations with respect to the structure of the fluid atoms. We wish to point out, however, that this is without loss of generality, since each non well-posed model can be transformed into a well-posed one \emph{without} changing the underlying ODE system, see \cite{efl_tcs}.

\begin{definition}[Well-posedness]\label{def_well_posedness}
A FEPA model $M$ is said to be well-posed if and only if for all occurrences $M_1 \parallel_{L} M_2$ in $M$ it holds that $\exists V_1 \big( \arate(M_1,V_1) > 0 \big) \ \land \ \exists V_2 \big( \arate(M_2,V_2) > 0 \big)$ for all $\alpha \in L$.
\end{definition}
In essence, a model is well-posed whenever any synchronised action may be performed by both operands, for some population function. Thus, $\pepa{Sys}$ is well-posed, but $P_1 \parallel_{\{\beta\}} Q$ is not since $Q$ does not do $\beta$-actions.

\section{Exact Aggregations}

\paragraph{Exact Fluid Lumpability.}
As discussed in Section~\ref{sec:intro}, EFL reduces the ODE system size by exploiting the fact that distinct fluid atoms with the same initial population function may have undistinguishable ODE solutions.

\begin{definition}[Exact Fluid Lumpability (EFL)]\label{def_exact_fluid_lumpability}
Let $M$ be a FEPA model and $\{\mathcal{P}^1, \ldots, \mathcal{P}^n\}$, where $\mathcal{P}^i = \{ P^i_j \mid 1 \leq j \leq k_i \}$, be a partition of $\mathcal{G}(M)$. The partition is called \emph{exactly fluid lumpable} if there exist bijections
\[ \sigma_{P^i_j} : ds(P^i_1) \rightarrow ds(P^i_j), \qquad 1 \leq i \leq n, \ 1 \leq j \leq k_i, \]
where $\sigma_{P^i_1} \equiv \text{id}_{ds(P^i_1)}$, such that for all initial populations $V(0)$ which satisfy
\[ V_P(0) = V_{\sigma_{P^i_j}(P)}(0), \qquad \forall 1 \leq i \leq n \forall P \in ds(P^i_1) \forall 1 \leq j \leq k_i \]
the same holds for all $t \geq 0$ in the corresponding ODE solution $V$, i.e.
\[ V_P(t) = V_{\sigma_{P^i_j}(P)}(t), \qquad \forall 1 \leq i \leq n \forall P \in ds(P^i_1) \forall 1 \leq j \leq k_i \forall t \geq 0 . \]
\end{definition}
Exact fluid lumpability of a partition is induced by the notion of label equivalence, established between tuples of fluid atoms. Intuitively, relating two tuples (of the same length) $(S_1, T_1)$ and $(S_2, T_2)$ means that, component-wise, the fluid atoms have the same trajectories; that is, $S_1$ (resp., $T_1$), has the same ODE solution as $S_2$ (resp., $T_2$). Fluid atoms within the same tuple are coupled in the sense that matching fluid atoms have to be provided for each element of a tuple.

\begin{definition}[Label Equivalence]\label{def_label_eq}
Let $M$ be a FEPA model and let $\mathcal{P} = (\vec{P}^1, \ldots, \vec{P}^N)$, $\vec{P}^i = (P^i_1, \ldots, P^i_{K_i})$, be a \emph{tuple partition} on $\mathcal{G}(M)$, that is, for each $P \in \mathcal{G}(M)$ there exist unique $1 \leq i \leq N$ and $1 \leq k \leq K_i$ with $P = P^i_k$. $\vec{P}^i$ and $\vec{P}^j$ are said to be \emph{label equivalent}, written $\vec{P}^i \sim_\mathcal{P} \vec{P}^j$, if $K_i = K_j$ and there exist bijections $\sigma_{k}: ds(P^i_k) \rightarrow ds(P^j_k)$, where $1 \leq k \leq K_i$, such that for all population functions $V$ of $M$ and
\[
V^\sigma_P :=
\begin{cases}
    V_{\sigma_k(P)} & \ , \exists 1 \leq k \leq K_i ( P \in ds(P^i_k) ) \\
    V_{\sigma_k^{-1}(P)} & \ , \exists 1 \leq k \leq K_i ( P \in ds(P^j_k) ) \\
    V_{P} & \ , \text{otherwise}
\end{cases}
\]
it holds that
\begin{enumerate}[a)]
    \item $\crate(M, V, P) = \crate(M, V^\sigma, \sigma_k(P))$
   	\item $\dsl \sum_{P'} p_\alpha(P', P) \crate(M, V, P') = \sum_{P'} p_\alpha(P', \sigma_{k}(P)) \crate(M, V^\sigma, P')$
   	\item $\crate(M, V, P) = \crate(M, V^\sigma, P)$ for all $P \in ds(P^l_k)$ with $P^l_k \notin \vec{P}^i,\vec{P}^j$
	\item $\arate(M,V) = \arate(M,V^\sigma)$ and $\arate(P^i_k) = \arate(P^j_k)$ for all $1 \leq k \leq K$.
\end{enumerate}
\end{definition}

\begin{example}\label{ex:second}
EFL has been used to simplify replicas of composite processes~\cite{Tschaikowski2013a}. For instance, let us consider the fluid atoms in Example~\ref{ex:first}, $R_d \rmdef (\alpha, \tilde{r}).R'_d$, and $R'_d \rmdef (\gamma, \tilde{s}).R_d$. Further, let us take the FEPA model
$$\pepa{Sys}_\mathcal{E} := \big((P_1 \parallel_{\{\alpha\}} R_1) \parallel_\emptyset \ldots \parallel_\emptyset (P_D \parallel_{\{\alpha\}} R_D)\big) \parallel_{\{\alpha\}} Q \ ,$$
which features $D$ replicas of composite processes of type $P_d \parallel_{\{\alpha\}} R_d$.  Let us consider now the tuple partition $\mathcal{P} := \{(P_1,R_1), \ldots,  (P_D,R_D), (Q)\}$. Then, it can be shown that $(P_i,R_i) \sim_\mathcal{P} (P_j,R_j)$, thus formalising the intuitive idea that each replica has the same solution (if initialised with identical conditions).
\end{example}

Using label equivalence, which acts on \emph{tuples} of labels, we define the following notion which allows to relate single labels.

\begin{definition}[Projected Label Equivalence]\label{def_proj_label_eq}
Fix a FEPA model $M$ and a tuple partition $\mathcal{P}$ of $\mathcal{G}(M)$. Two fluid atoms $P_1,P_2 \in \mathcal{G}(M)$ are projected label equivalent, $P_1 \approx_\mathcal{P} P_2$, if $\vec{P}^i \sim_\mathcal{P} \vec{P}^j$ and $k_i = k_j$ in the unique assignment $P_1 = P_{k_i}^i$, $P_2 = P_{k_j}^j$.
\end{definition}
Therefore, we have that $P_1 \approx_{\cal P} P_2$, $R_1 \approx_{\cal P} R_2$, and so on.

Following~\cite{concur12}, we are finally ready to extend EFL to FEPA, showing that the following is valid also for semantics based on the law of mass action, $\rho = \cdot$.%

\begin{theorem}\label{thm_proj_label_eq}
Fix a FEPA model $M$ and a tuple partition $\mathcal{P}$ of $\mathcal{G}(M)$. Then, $\sim_\mathcal{P}$ is a congruence relation with respect to $\parallel_L$, $\approx_\mathcal{P}$ is an equivalence relation on $\mathcal{G}(M)$ and $\mathcal{G}(M) / \approx_\mathcal{P}$ is exactly fluid lumpable.
\end{theorem}
For instance, it holds that $\mathcal{G}(\pepa{Sys}_\mathcal{E}) / \approx_\mathcal{P}$ yields the exactly fluid lumpable partition
$\big\{ \{P_1, \ldots, P_D\}, $ $ \{R_1, \ldots, R_D\}, \{Q\} \big\}$.
Let us notice that this result gives us a tool which aggregates ODE systems to smaller ones. In Example~\ref{ex:second}, for instance, this allows one to recover the solution of an ODE system of size $4D + 2$ by solving an aggregated ODE system of size $4 + 2$.

Furthermore, the characterisation of EFL can be shown also when $\rho = \cdot$.

\begin{theorem}\label{thm_char_efl}
Fix a well-posed FEPA model $M$, a tuple partition $\mathcal{P} = \{\vec{P}^1, \ldots$ $\ldots, \vec{P}^N\}$ on $\mathcal{G}(M)$ and assume that $\vec{P}^i \sim_\mathcal{P} \vec{P}^j$. Then, $P^i_k$ is semi-isomorphic to $P^j_k$ for all $1 \leq k \leq K_i$.
\end{theorem}

Using the last theorem, one can show that, under the condition of well-posedness, different exactly fluid lumpable partitions can be merged.

\begin{theorem}\label{thm_merge}
Fix a well-posed FEPA model $M$ and a set of tuple partitions $S = \{\mathcal{P}_1, \ldots, \mathcal{P}_m\}$ of $\mathcal{G}(M)$. Then, the partition $\mathcal{G}(M) / (\approx_{\mathcal{P}_1} \cup \ldots \cup \approx_{\mathcal{P}_m})^*$ is exactly fluid lumpable.
\end{theorem}

{
\bfseries
\begin{remark}
\textnormal{In that what follows, we will assume that an EFL partition is established as in Theorem \ref{thm_proj_label_eq}.
}\end{remark}
}

\paragraph{Ordinary Fluid Lumpability.} EFL considers a partition of fluid atoms such that elements in the same part have the \emph{same solution}. Instead, in ordinary fluid lumpability (OFL) \emph{the sum of the solutions} of elements within the same part are fully recovered from the solution of a (smaller) ODE system consisting of one single ODE for a representative element of each part.

\begin{definition}[Ordinary Fluid Lumpability (OFL)]\label{def_ordfluidlump}
Let $M$ be a FEPA model and let $\{\mathcal{P}^1, \ldots, \mathcal{P}^n\}$ be a partition of $\mathcal{G}(M)$. The partition is called \emph{ordinarily fluid lumpable} if there exist bijections
\[ \sigma_{P^i_j} : ds(P^i_1) \rightarrow ds(P^i_j), \qquad 1 \leq i \leq n, \ 1 \leq j \leq k_i \]
such that $\sigma_{P^i_1} \equiv \text{id}_{ds(P^i_1)}$ and for all $\alpha \in \mathcal{A}$, $1 \leq i \leq n$ and $V$, it holds that
\begin{enumerate}[i)]
	\item $\dsl \sum_{1 \leq j \leq k_i} \crate(M,V,\sigma_{P^i_j}(P)) = \crate(M,V^\sigma,P), \quad \forall P \in ds(P^i_1)$
    \item $\dsl \sum_{1 \leq j \leq k_i} \sum_{P' \in ds(P^i_1)} p_\alpha(\sigma_{P^i_j}(P'), \sigma_{P^i_j}(P)) \crate(M,V,\sigma_{P^i_j}(P'))$ \vspace{-0.25cm}
    \[ \quad = \sum_{P' \in ds(P^i_1)} p_\alpha(P', P) \crate(M,V^\sigma,P'), \quad \forall P \in ds(P^i_1) \]
	\item $\arate(M,V) = \arate(M,V^\sigma)$ and $\arate(P^i_1) = \arate(P^i_j)$ for all $1 \leq j \leq k_i$,
\end{enumerate}
where $V^\sigma_P :=
\begin{cases}
    \dsl \sum_{1 \leq j \leq k_i} V_{\sigma_{P^i_j}(P)} & \ , \exists 1 \leq i \leq n \big( P \in ds(P^i_1) \big) \\
    0 & \ , \text{otherwise}.
\end{cases}$
\end{definition}

We can now define and relate the lumped ODE system to the original one.

\begin{theorem}[ODE Lumping]\label{thm_ode_lump}
Let $M$ be a FEPA model, $\{\mathcal{P}^1, \ldots, \mathcal{P}^n\}$ an ordinarily fluid lumpable partition of $\mathcal{G}(M)$, and $V$ the ODE solution of $M$ for a given initial condition $V(0)$. Define
\[ W_P := \sum_{1 \leq j \leq k_i} V_{\sigma_{P^i_j}(P)}, \qquad 1 \leq i \leq n, \quad P \in ds(P^i_1)   \]
and
\[ \overline{W}_P :=
\begin{cases}
W_P & , \ \exists 1 \leq i \leq n \big( P \in ds(P^i_1) \big) \\
0 & , \ \text{otherwise}
\end{cases} \]
for all $P \in \mathcal{B}(M)$. Then, $W$ is the unique solution of the ODE system
\begin{align}\label{thm_ord_lump_general}
\dot{W}_P & = \sum_{\alpha \in \mathcal{A}} \Big(\sum_{P' \in ds(P^i_1)} p_\alpha(P',P) \crate(M,\overline{W},P') - \crate(M,\overline{W},P) \Big), \nonumber \\
W_P(0) & = \sum_{1 \leq j \leq k_i} V_{\sigma_{P^i_j}(P)}(0),
\end{align}
where $1 \leq i \leq n$ and $P \in ds(P^i_1)$. Hence,
\[ \sum_{1 \leq j \leq k_i} V_{\sigma_{P^i_j}(P)}, \qquad 1 \leq i \leq n, \quad P \in ds(P^i_1), \]
can be recovered by solving the lumped ODE system (\ref{thm_ord_lump_general}).
\end{theorem}
For instance, let us consider again Example~\ref{ex:first}. It can be shown that the partition $\big\{ \{P_1, \ldots, P_D\}, \{Q\} \big\}$ is an OFL partition of $\pepa{Sys}$. According to the above theorem, the aggregated ODE system, of size 4, is given by
\begin{align}
\dot{W}_{P} & = - ru W_P \cdot W_Q + s W_{P'} &
\dot{W}_{P'} & = + ru W_P \cdot W_Q - s W_{P'} \nonumber \\
\dot{W}_{Q} & = - ru W_P \cdot W_Q + w W_{Q'} &
\dot{W}_{Q'} & = + ru W_P \cdot W_Q - w W_{Q'} \nonumber 
\end{align}
with initial condition $W_P(0) = \sum_{d = 1}^D V_{P_d}(0) =\sum_{d = 1}^D N_d$, $W_{P'}(0) = \sum_{d = 1}^D V_{P'_d}(0)$  $= 0$, $W_Q(0) = V_Q(0) = M$, and $W_{Q'}(0) = V_{Q'}(0) = 0$. It holds that $W_P(t) = \sum_{d = 1}^D V_{P_d}(t)$, but each individual solution, $V_{P_d}(t)$,  cannot be recovered.

The next theorem states the congruence property of OFL with respect to the parallel composition of FEPA.

\begin{theorem}[Congruence]\label{thm_cong}
Let us fix two FEPA models $M_1, M_2$ and assume that $\{\mathcal{P}^1, \ldots, \mathcal{P}^n\}$ and $\{\mathcal{P}^{n+1}, \ldots, \mathcal{P}^{n+m+1}\}$ are ordinarily fluid lumpable partitions of $\mathcal{G}(M_1)$ and $\mathcal{G}(M_2)$, respectively. Then, thanks to the set of bijections
\[ \sigma_{P^i_j} : ds(P^i_1) \rightarrow ds(P^i_j), \qquad 1 \leq i \leq n+m+1, \quad 1 \leq j \leq k_i , \]
the partition $\{\mathcal{P}^1, \ldots, \mathcal{P}^n\} \cup \{\mathcal{P}^{n+1}, \ldots, \mathcal{P}^{n+m+1}\}$ of $\mathcal{G}(M_1 \parallel_{L} M_2)$ is also ordinarily fluid lumpable.
\end{theorem}

Finally, similarly to EFL, the next theorem characterises OFL with respect to semi-isomorphism.

\begin{theorem}\label{thm_char_ofl}
Fix a well-posed FEPA model $M$ and assume that the partition $\{\mathcal{P}^1, \ldots, \mathcal{P}^n\}$ of $\mathcal{G}(M)$ is ordinarily fluid lumpable. Then, $P^i_1$ is semi-isomorphic to $P^i_j$ for all $2 \leq j \leq k_i$ and $1 \leq i \leq n$.
\end{theorem}
Well-posedness is needed in Theorem \ref{thm_char_efl} and \ref{thm_char_ofl}. Let $\tilde{P}_d \rmdef (\alpha,r).\tilde{P}'_d + (\gamma,r/d).\tilde{P}_d + (\gamma,r - r/d).\tilde{P}_d$ and $\tilde{P}_d \rmdef (\beta,s).\tilde{P}'_d$, i.e. $\tilde{P}_i$ and $\tilde{P}_j$ are not semi-isomorphic. Then, $\big((\tilde{P}_1 \parallel_{\{\alpha\}} R_1) \parallel_\emptyset \ldots \parallel_\emptyset (\tilde{P}_D \parallel_{\{\alpha\}} R_D)\big) \parallel_{\{\alpha,\gamma\}} Q$ and $(\tilde{P}_1 \parallel_{\emptyset} \ldots \parallel_{\emptyset} \tilde{P}_D) \parallel_{\{\alpha,\gamma\}} Q$ are ill-posed, while $(\tilde{P}_i,R_i) \sim_{\mathcal{P}} (\tilde{P}_j,R_j)$ and $\big\{ \{ \tilde{P}_1, \ldots, \tilde{P}_D \}, \{Q\} \big\}$ is lumpable.



\section{Fluid $\varepsilon$-Lumpability}\label{sec:theory:epsilon}

We now study aggregations for models where certain fluid atoms can be made elements of the same partition block after a suitable perturbation of their parameters. In the case of EFL, we allow different rates and initial populations. In the case of OFL, instead, we only consider the former because there is no requirement on identical initial populations for aggregated fluid atoms.

At the basis of our investigation is the following comparison theorem which relates two ODE systems of the same size, where the vector field is made dependent on a vector of parameters, here denoted by $\zeta$ and $\xi$. Thus, for some norm $\norm{\cdot}$, we interpret $\varepsilon = \norm{\xi - \zeta}$ as the intensity of the perturbation on the rates of the same model. The two ODE systems may also have different initial conditions $\underline{x}_\zeta$ and $\underline{x}_\xi$, and we let $\delta  = \norm{\underline{x}_\xi - \underline{x}_\zeta}$. This will be used to define our approximate version of EFL. The theorem states that, on a fixed time interval $[0;t]$, the distance between the two solutions depends \emph{linearly} on both $\varepsilon$ and $\delta$.
\begin{theorem}\label{thm_peturbation_0}
Consider the ODE systems
\begin{align*}
\begin{cases}
\dot x_\zeta = f(\zeta,x_\zeta) \\
x_\zeta(0)= \underline{x}_\zeta
\end{cases}
\qquad\qquad\qquad
\begin{cases}
\dot x_\xi = f(\xi,x_\xi)\\
x_\xi(0)= \underline{x}_\xi
\end{cases}
\end{align*}
where $f$ is assumed to be Lipschitz continuous in some domain $\mathcal{D} \subseteq \xR^{n+m}$, both with respect to $x$ as with respect
to $\zeta$ with Lipschitz constant $L_\zeta$ and $K_x$ respectively, that is
\begin{align*}
\norm{f(\zeta,x)-f(\zeta,x')} & \leq L_\zeta \norm{x-x'} , \qquad (\zeta,x), (\zeta,x')\in \mathcal{D}, \\
\norm{f(\zeta,x)-f(\zeta',x)} & \leq K_x \norm{\zeta-\zeta'} , \qquad (\zeta,x), (\zeta',x)\in \mathcal{D}.
\end{align*}
Let us assume further that both ODE systems have a solution on $[0;t]$, where $t > 0$, which remains in $\mathcal{D}$, and that $K := \sup_{0 \leq s \leq t} K_{x_\xi(s)} < \infinity$. Then
\[
\norm{x_\zeta(t)-x_\xi(t)}\leq \left(\frac{\varepsilon K}{L_\zeta} + \delta\right)\xe^{L_\zeta t} - \frac{\varepsilon K}{L_\zeta}
\]
if $\varepsilon = \norm{\xi-\zeta}$ and $\delta = \norm{\underline{x}_\zeta - \underline{x}_\xi}$.
\end{theorem}

Next, we formally introduce the notion of perturbation on rates.


\begin{definition}
For a FEPA model $M$, let $\nu(M)$ denote the vector of distinct occurrences of action rates in $M$, written $\nu(M) = (x_1, \ldots, x_i, \ldots, x_{\lvert \nu(M) \rvert})$. Then, for a $\xi \in \mathbb{R}_{>0}^{|\nu(M)|}$, the model $M(\xi)$ arises from $M$ by replacing each $x_i$ with $\xi_i$.
\end{definition}

Theorem~\ref{thm_peturbation_0} can be applied to FEPA  by proving that FEPA models induce globally Lipschitz ODE systems and have bounded ODE solutions.

\begin{theorem}\label{thm_perturbation}
Fix a FEPA model $M$, a $\zeta \in \mathbb{R}_{>0}^{|\nu(M)|}$, a population function $V^\zeta(0)$ and $c, t > 0$. Then, there exist $C_1, C_2 > 0$ such that $\| V^\xi(0) \|, \| \xi \| \leq c$ implies
\[ \max_{0 \leq s \leq t} \| V^\xi(s) - V^\zeta(s) \| \leq C_1 \| \xi - \zeta \| + C_2 \| V^\xi(0) - V^\zeta(0) \| , \]
where $V^\xi$ and $V^\zeta$ refer to the ODE solutions of $M(\xi)$ and $M(\zeta)$, respectively.
\end{theorem}
Let us remark that the above result states that the perturbations on the rate parameters and on the initial conditions yield separate additive contributions to the aggregation error. While the former kind of perturbation can be related to analogous approximate aggregations for Markov chains, e.g.,~\cite{BuchholzOrdinaryExact}, the latter does not have a stochastic counterpart to the best of our knowledge. This is because altering the initial population of agents leads to a generator matrix of different size, while a perturbation on the rates preserves the matrix structure. In the fluid semantics, instead, both cases do not alter the structure of the ODE system, but only its parameters.
Let us also notice that, in the above theorem, $M(\zeta)$ is arbitrary but fixed, whereas $M(\xi)$ varies. We now focus  on the situation where $M(\zeta)$ has either an exactly or an ordinarily fluid lumpable partition.

\begin{definition}[Fluid $\varepsilon$-Lumpability]
Fix a FEPA model $M$ and $\xi \in \mathbb{R}_{>0}^{|\nu(M)|}$. If $M(\zeta)$ has an exactly/ordinarily fluid lumpable partition $\{\mathcal{P}^1, \ldots, \mathcal{P}^n\}$ for some $\zeta \in \mathbb{R}_{>0}^{|\nu(M)|}$, $M(\xi)$ is said to be $\norm{\xi - \zeta}$-exactly/ordinarily fluid lumpable with respect to some norm $\norm{\cdot}$.
\end{definition}
For instance, let us take $D = 2$ and, with obvious ordering of the rate occurrences, $\zeta = (r,s,r,s, u,w)$, for which (\ref{eq_running_example_fpa}) admits EFL whenever $V_{P_1} = V_{P_2}$ and $V_{P'_1} = V_{P'_2}$. Consider now the same model, where the rates are replaced with $\xi = (r + \varepsilon_1, s + \varepsilon_2, r, s, u, w)$. This model is $\varepsilon$-exactly fluid lumpable with $\varepsilon = \norm{(\varepsilon_1, \varepsilon_2, 0, 0, 0, 0)}$.
In general, an exactly/ordinarily fluid lumpable partition admits an infinity of $\varepsilon$-lumpable partitions; $\varepsilon$ gives the measure of how close a given model is to error-free lumping.

Both $\varepsilon$-EFL and $\varepsilon$-OFL enjoy congruence.
\begin{theorem}[Congruence]\label{thm_eps_cong}
Fix two FEPA models $M_1, M_2$ and assume that $\{\mathcal{P}^1_k, \ldots, \mathcal{P}^{n_k}_k\}$ is $\norm{\xi_k - \zeta_k}$-exactly/ordinarily fluid lumpable in $\mathcal{G}(M_k(\xi_k))$ for some $\xi_k,\zeta_k \in \mathbb{R}_{>0}^{|\nu(M_k)|}$, and $k = 1,2$. Then, for any $L \subseteq \mathcal{A}$, $\bigcup_{k=1}^2 \{\mathcal{P}^1_k, \ldots, \mathcal{P}^{n_k}_k\}$ is $\norm{(\xi_1,\xi_2) - (\zeta_1,\zeta_2)}$-exactly/ordinarily fluid lumpable in $\mathcal{G}(M_1 \parallel_{L} M_2)$.
\end{theorem}
Clearly, as an OFL partition does not depend on the initial values, a perturbation of initial values is interesting only in the case of EFL.

We turn now to a characterisation of $\varepsilon$-OFL and $\varepsilon$-EFL. It is natural consider an $\varepsilon$-extension of semi-isomorphism to relate fluid atoms that are isomorphic up to an appropriate perturbation of their rates.

\begin{definition}[$\varepsilon$-Semi-Isomorphism]
Two fluid atoms $P$ and $Q$ are $\varepsilon$-semi-isomorphic for some $\varepsilon > 0$, if there is a bijection $\sigma : ds(P) \rightarrow ds(Q)$ which satisfies
\[ \Big| \sum_{P_i \xrightarrow{(\alpha, r)} P_j} r - \sum_{\sigma(P_i) \xrightarrow{(\alpha, r)} \sigma(P_j)} r \Big| \leq \varepsilon\]
for all $P_i, P_j \in ds(P)$ and $\alpha \in \mathcal{A}$. Such $\sigma$ is called $\varepsilon$-semi-isomorphism.
\end{definition}

Analogously to EFL and OFL, the following characterises $\varepsilon$-lumpability with respect to $\varepsilon$-semi-isomorphism.
\begin{theorem}\label{thm_eps_char}
For any well-posed FEPA model $M$ and norm $\norm{\cdot}$, there exists a $C > 0$ such that if $\{\mathcal{P}^1, \ldots, \mathcal{P}^n\}$ is an $\norm{\xi-\zeta}$-exactly/ordinarily fluid lumpable partition of $M(\xi)$, where $\xi, \zeta \in \mathbb{R}_{>0}^{|\nu(M)|}$, then:
\begin{enumerate}[1)]
   \item Any two fluid atoms $P^i_j, P^i_{j'}$ of $M(\xi)$ are $C \norm{\xi-\zeta}$-semi-isomorphic;
   \item In the special case where for all $\alpha \in \mathcal{A}$ and $P,P' \in \mathcal{B}(M)$ there is at most one $\alpha$-transition from $P$ to $P'$ and $\norm{\cdot} = \norm{\cdot}_\infinity$, $1)$ holds for $C = 1$.
\end{enumerate}
\end{theorem}
For instance, the above theorem ensures that $P_d$, $P_{d'}$ are $\norm{\xi-\zeta}_\infinity$-semi-isomorphic in $\pepa{Sys}(\xi)$, cf. (\ref{eq_running_example_fpa}), for all $\xi \in \mathbb{R}_{>0}^{|\nu(\pepa{Sys})|}$, if $\zeta = (r, s, \ldots,r,s,r,w)$.

%
\section{Numerical Evaluation}

\begin{figure}[t]
\subfloat[$\varepsilon$-EFL for $\rho = \min$]{
\label{fig:efl.min}
\centering
\includegraphics[scale=0.64]{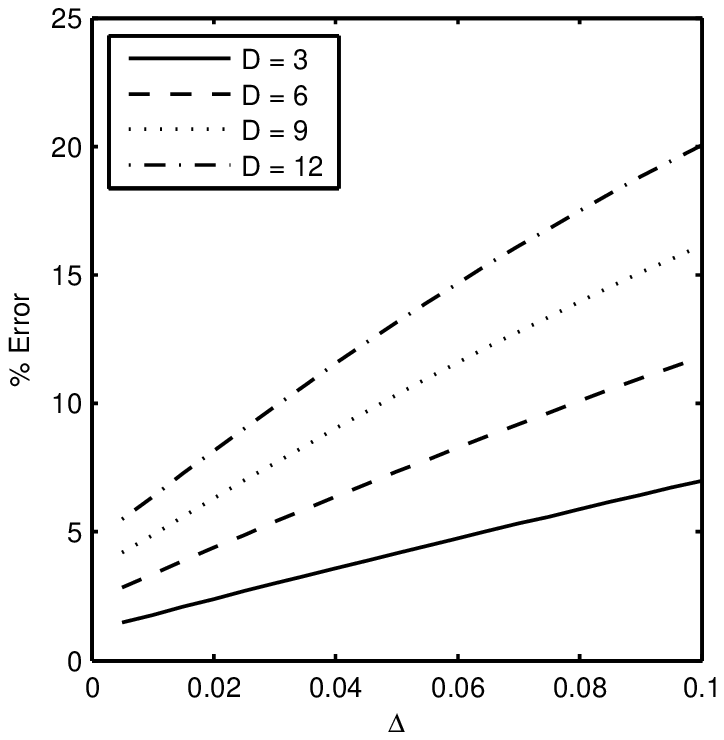}
}
\subfloat[$\varepsilon$-EFL for $\rho = \cdot$]{
\label{fig:efl.prod}
\centering
\includegraphics[scale=0.64]{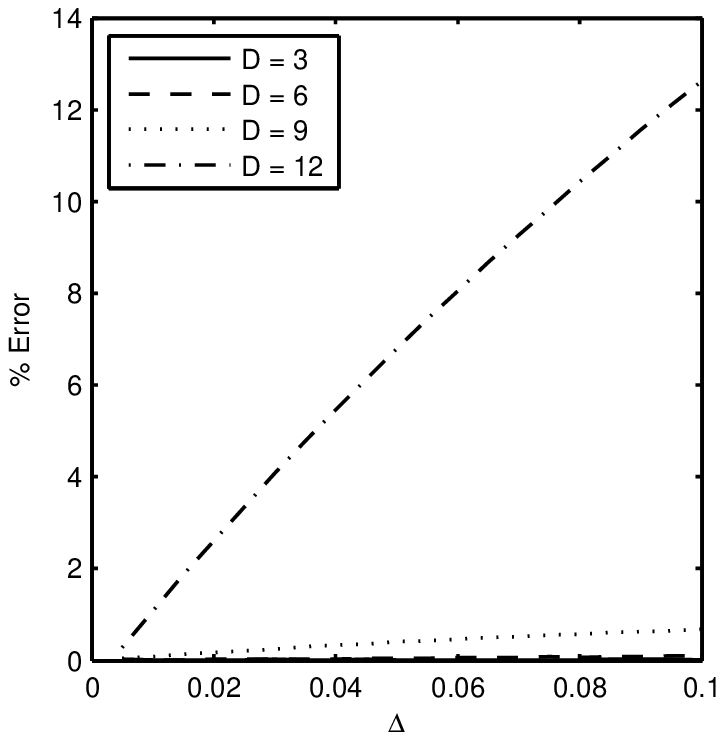}
}
\subfloat[$\varepsilon$-OFL, for $D = 12$]{
\label{fig:ofl}
\centering
\includegraphics[scale=0.64]{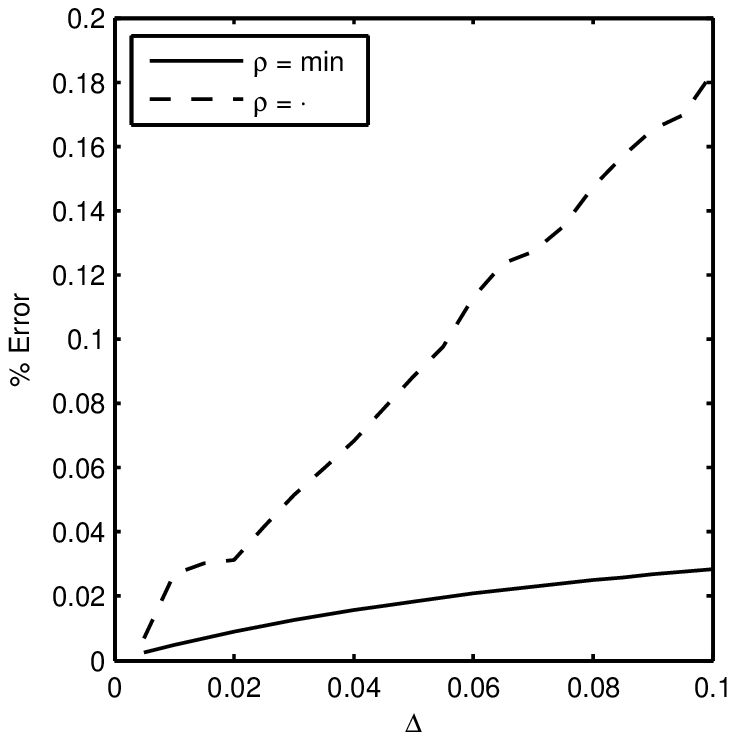}
}
\caption{Numerical evaluation of $\varepsilon$-lumpability.}\label{fig:elf}
\end{figure}
In this section we provide some numerical evidence of the aggregation error introduced by $\varepsilon$-EFL and $\varepsilon$-OFL\@. We considered the FEPA model (\ref{eq_running_example_fpa}) with
\begin{equation}\label{eq:original}
P_d \rmdef (\alpha, r_d).P'_d \qquad P'_d \rmdef (\beta, s).P_d \qquad Q = (\alpha, r).Q' \qquad Q' = (\gamma, w).Q \ .
\end{equation}
We fixed $s = 0.5$, $w = 15.0$. To obtain non-isomorphic fluid atoms, we set $r_d =  1.0 + (d-1)\Delta$, where $\Delta$ is a parameter that was varied  between 0.0005 and 0.1000 at 0.005 steps in our tests. In this way, $\Delta$ is related to the intensity of the perturbation.
The initial population function was set as
$V_{P_d}(0) = 200 + (d - 1)$, $V_{P'_d}(0) = 0$, $V_Q(0) = 400$, and $V_{Q'}(0) = 0$; thus, the $P_d$ fluid atoms  have initial populations separated by a few percent.
For evaluating both $\varepsilon$-EFL and $\varepsilon$-OFL, we considered a perturbed model where $r_d$ in (\ref{eq:original}) was made independent of $d$ and set equal to the average value in the original model, i.e., $1.0 + (\Delta/D) \sum_{d = 1}^D (d-1)$. In such a perturbed model, all $P_d$ fluid atoms are now isomorphic.
\paragraph*{Assessment of $\varepsilon$-EFL\@.}  We considered different values of $D$ to numerically evaluate the impact of different initial conditions on the quality of the aggregation of $\varepsilon$-EFL. Specifically, we set $D = 3,6,9,12$. Let us recall that (\ref{eq_running_example_fpa}) has $2D + 2$ ODEs. For each value of $D$ and $\Delta$, the model solution was compared against that of the perturbed model with the initial population function set as follows:
$V^\varepsilon_{P_d}(0) = 200 + (1/D) \sum_{d=1}^D (d-1)$, $V^\varepsilon_{P'_d}(0) = 0$, $V^\varepsilon_Q(0) = 400$, and $V^\varepsilon_{Q'} = 0$.
In this way, the initial population of $P_d$ fluid atoms is made independent from $d$ and is set equal to the average initial population across $d$, similarly to what done for the perturbation on $r_d$. It follows that, in the perturbed model, $\{ \{ P_1, \ldots, P_D \}, \{ Q \} \}$ is an exactly fluid lumpable partition. Hence, the original model and the perturbed one are related by $\varepsilon$-EFL.
Both models were solved over the time interval $[0;100]$ (ensuring convergence to equilibrium in all cases), with solutions registered at 0.2 time steps. The approximation relative error for $\varepsilon$-EFL is as:
\[
100 \times \max_{t \in \{ 0,0.02,\ldots, 100 \}} \max_{S \in \{ P_1, \ldots, P_d, Q \} }  \frac{\left \lvert V_S(t) - V_S^\varepsilon(t)\right \rvert}{V_{S}(0)},
\]
where $V_S(t)$ is the solution of the original model and $V_S^\varepsilon(t)$ is the corresponding solution in the  perturbed one. The absolute difference is normalised with respect to the total population of the fluid atom.

The results are presented in Figures~\ref{fig:efl.min} and \ref{fig:efl.prod}, for  $\rho = \min$ and $\rho = \cdot$, respectively. In both cases, we observe a linear growth of the error as a function of the perturbation $\Delta$. For any fixed $D$, the case $\rho = \cdot$ yields more accurate aggregates than $\rho = \min$, with particularly small errors for $D = 3,6,9$. These tests show that even non-negligible perturbations (i.e., up to $\Delta$ ca 0.04) can produce acceptable errors (i.e., less than 10\%) in practice.
\paragraph*{Assessment of $\varepsilon$-OFL\@.} A similar setting was used for the assessment of  $\varepsilon$-OFL, since in the perturbed model $\{ \{ P_1, \ldots, P_D \}, \{ Q \} \}$ is also an ordinarily fluid lumpable partition. However, unlike $\varepsilon$-EFL,  as discussed, $\varepsilon$-OFL \emph{does not} depend on the initial population function. Therefore, in our tests the initial conditions were not changed in the perturbed model. Furthermore, we analysed only the case $D = 12$, which yielded the worst accuracy in $\varepsilon$-EFL; the other cases showed the same errors (up to numerical precision of the ODE solver). A different error metric was used, to reflect the fact that OFL involves sums of ODE solutions of the unaggregated model. The approximation relative error is defined as:
\[
100 \times  \max_{t \in \{ 0,0.02,\ldots, 100 \}} \max \left \{ \frac{\left \lvert \sum_{d = 1}^D V_{P_d}(t) - V_P^\varepsilon(t)\right \rvert}{\sum_{d = 1}^D V_{P_d}(0)}, \frac{\lvert V_Q(t) - V^\varepsilon_Q(t) \rvert}{V_Q(0)} \right \} ,
\]
where $V_P^\varepsilon(t)$ and $V_Q^\varepsilon(t)$ are the solutions in the OFL model corresponding to the sum of the $P_d$ derivatives and to the $Q$ derivative, respectively.
The results are shown in Figure~\ref{fig:ofl}. Overall, both for $\rho = \min$ and $\rho = \cdot$, the $\varepsilon$-OFL appears to be much more robust, with negligible errors across all values of $\Delta$.

\section{Conclusion}
This paper has studied ODE aggregations for a process algebra that uniformly treats two different dynamics of interactions, for capturing models of performance and chemical reaction networks. Our approximate aggregations allow models with heterogeneous processes to be treated as homogeneous models by appropriate perturbations of rate parameters and initial populations. Although the numerical results suggest that this aggregation can be robust,  tightening of the theoretical error bound is part of future work for increasing the a-priori confidence on the practical usefulness of these techniques.

\paragraph{Acknowledgement.} This work has been partially supported by the EU project QUANTICOL, 600708, and by the DFG project FEMPA.

\bibliographystyle{eptcs}
\bibliography{fluid_equivalences}

\newpage

\end{document}